\documentstyle[12pt]{article}
\begin{document}
\begin{center}
{\large On Superconductors and Torsion Vortices}
\end{center}
\vspace{2cm}
\begin{center}
by L.C.Garcia de Andrade\footnote{Grupo de Relatividade e Teoria de Campo - 
Instituto de F\'{\i}sica - UERJ, Rio de janeiro, R.J 20550, Brasil} 
\end{center}
\begin{abstract}
\hspace{0.6cm}The Meissner effect for superconductors in spacetimes with torsion is revisited.Two new physical interpretaions are presented.The first considers the Landau-Ginzburg theory yields a new symmetry-breaking vacuum depending on torsion.In the second interpretation a gravitational Meissner torsional effect where when the Higgs field vanishes, torsion and electromagnetic fields need not vanish and outside the Abrikosov tubes a torsion vector analogous to the Maxwell potential is obtained.The analogy is even stronger if we think that in this case the torsion vector has to be derivable from a torsion potential.Another solution of Landau-Ginzburg equation is shown to lead naturally to the geometrization of the electromagnetism in terms of the torsion field.
\end{abstract}
\vspace{2cm}
\begin{center}
\end{center}
\newpage
\hspace{0.6cm}
Ealier D'Auria and Regge \cite{1} have considered gravity theories with torsion and gravitationally asymptotically flat instantons.The vanishing of the gap inside a flux tube had a perfect analogy in the vanishing of the vierbein inside the gravitational instanton.Connection between torsion vortices and non-trivial topological configurations have also been pointed out in \cite{2}.In the following an argument considering the analogy between the Meissner effect and vanishing torsion in gravity is revisited.We shown that two distinct physical interpretations may be given to the Landau-Ginzburg theory with torsion.In the first torsion shifts the symmetry breaking vacuum by a torsion dependent value.In the second the Landau-Ginzburg equation is solved and a torsion potential is obtained outside the Abrikosov tube in analogy with the Maxwell field derivable from a scalar electromagnetic potential in the usual Meissner effect.Finally an application is considered by computing the photon mass in the case of the magnetic field of a neutron star.Let us now considered the Landau-Ginzburg equation for superconductors
\begin{equation}
D_{\mu}D^{\mu}{\phi}=g{\phi}(|\phi|^{2}-{\lambda}^{2})
\label{1}
\end{equation}
where ${\phi}$ is the gap field and g is a coupling constant.
In the non-gravitational Meissner effect near the symmetry-breaking 
vacuum the magnitude of ${\phi}$ is constant
\begin{equation}
|{\phi}|={\lambda}
\label{2}
\end{equation}
By considering the extended covariant derivative to include 
torsion
\begin{equation}
D_{\mu}={\partial}_{\mu}-ieA_{\mu}-ifQ_{\mu}
\label{3}
\end{equation}
where $ Q_{\mu} $ are the components of the torsion vector.
Substitution of the (\ref{3}) into (\ref{1}) yields
\begin{equation}
{D^{em}}_{\mu}{D^{em}}^{\mu}{\phi}=g{\phi}{|\phi|^{2}-({\lambda}^{2}+\frac{f}{g}Q^{2})}
\label{4}
\end{equation}
Here $D^{em}$ is the electromagnetic part of the covariant 
derivative given in (\ref{3}).We also have made use of the 
following gauges $D^{em}_{\mu}Q^{\mu}=0$ and $Q^{\mu}D^{em}_{\mu}{\phi}=0$.
Thus it is easy to see from this equation that the 
symmetry-breaking vacuum is shift by a torsion energy term 
$Q^{2}=Q_{\mu}Q^{\mu}$.Situations like that may certainly appear 
in other problems in field theory like domain walls with torsion 
\cite{3,4} and problems in Superfluids \cite{5}.Let us now solve 
the equation (\ref{4}).Suppose that we work in the static case 
where we are from the wall and impurities and the general solution drifts into an asymptotic regime in which 
it is covariantly constant
\begin{equation}
D_{\mu}{\phi}={\partial_{\mu}-ieA_{\mu}-ifQ_{\mu}}{\phi}=0
\label{5}
\end{equation}
Differentiating once again we have
\begin{equation}
{D_{\mu}D_{\nu}-D_{\nu}D_{\mu}}{\phi}=ieF_{\mu \nu}{\phi}
-if\partial_{\mu}Q_{\nu}-\partial_{\nu}Q_{\mu}{\phi}=0
\label{6}
\end{equation}
Thus the vanishing of the Maxwell tensor is not necessarily followed 
by the vanishing of the torsion.Note that equation (\ref{6}) has two possible solutions.In the first case assuming that the space is free of electromagnetic fields or $F_{\mu\nu}=0$ the vanishing ofthe second term implies a vortice term given by
\begin{equation}
\partial_{\mu}Q_{\nu}-\partial_{\nu}Q_{\mu}=0
\label{7}
\end{equation}
nevertheless in places of the superconductor where the Higgs field
${\phi}$ vanishes,torsion and electromagnetic fields do not vanish 
as happens on the inside of the Abrikosov tubes.Outside the tube 
equation (\ref{7}) is obeyed and torsion is derived from a torsion 
potential in analogy to the electromagnetic potential
\begin{equation}
A_{\mu}=-\frac{i}{e}\partial_{\mu}\theta
\label{8}
\end{equation}
where the phase ${\theta}$ appears in the solution as 
\begin{equation}
\phi=\lambda e^{i\theta}
\label{9}
\end{equation}
By analogy torsion is given by
\begin{equation}
Q_{\mu}=\frac{i}{f}{\partial}_{\mu}{\alpha}
\label{10}
\end{equation}
where ${\alpha}$ is the torsion phase.To perform the above 
computations we have considered that second order terms in the 
coupling of torsion and the electromagnetic potential may be 
dropped.The second solution assumes that the space is not free of electromagnetic fields and equation (\ref{6}) implies that the electromagnetic fields can be written in terms of torsion as
\begin{equation}
F_{\mu\nu}=\frac{f}{e}\partial_{\mu}Q_{\nu}-\partial_{\nu}Q_{\mu}
\label{11}
\end{equation}
This implies that the electromagnetic vector potential is proportional to the torsion vector.Similar results have been obtained by R.Hammond \cite{6} in the more general case of the non-Riemannian geometry with curvature and torsion.Let us now consider a non-linear electrodynamics in 
spaces with torsion
\begin{equation}
L\cong\sqrt{-g}[R(\Gamma)(1+A^{2})-\frac{1}{4}F_{ij}F^{ij} + 
J^{i}A_{i}]
\label{13}
\end{equation}
The interaction between the vector potencial and background 
torsion was found to break the gauge invariance leading to a 
Proca type mass term of the form :
\begin{equation}
{m^{2}}_{\gamma}\cong\lambda R(\Gamma)A^{2}
\label{14}
\end{equation}
This term gives the photon a small rest mass.

\end{document}